\documentclass[12pt]{iopart}
\usepackage{iopams}
\usepackage{graphicx}

\begin{document}

\title[Detuned Mechanical Parametric Amplification as a QND Measurement]{Detuned Mechanical Parametric Amplification as a Quantum Non-Demolition Measurement}
\author{A Szorkovszky$^1$, AA Clerk$^2$, AC Doherty$^3$, and WP Bowen$^1$}
\address{$^1$ Centre for Engineered Quantum Systems, University of Queensland, St Lucia, Australia}
\address{$^2$ Department of Physics, McGill University, Montr\'eal, Canada}
\address{$^3$ Centre for Engineered Quantum Systems, University of Sydney, Sydney, Australia}
\ead{alexs@physics.uq.edu.au}

\begin{abstract}

Recently it has been demonstrated that the combination of continuous position detection with detuned parametric driving can lead to 
significant steady-state mechanical squeezing, far beyond the 3 dB limit normally associated with parametric driving.  In this work, we show the close connection between this detuned scheme and quantum non-demolition (QND) measurement of a single mechanical quadrature.  In particular, we show that applying an experimentally realistic detuned parametric drive to a cavity optomechanical system allows one to effectively realize a QND measurement despite being in the bad-cavity limit.  In the limit of strong squeezing, we show that this scheme offers significant advantages over standard backaction evasion, not only by allowing operation in the weak measurement and low efficiency regimes, but also in terms of the purity of the mechanical state.

\end{abstract}

\section{Introduction}

%Quantum measurement of oscillators
Recently, much attention has been focused on the problem of measuring a macroscopic harmonic oscillator at the level of its quantum mechanical fluctuations, and thereby controlling its quantum state. For example, measuring the position with an uncertainty smaller than the quantum zero-point motion results in a ``squeezed'' state. Such quantum states are the basis for sensing at an unprecedented scale\cite{caves} and for new kinds of information processing\cite{braunstein}. Squeezing the quantum noise in mechanical oscillators is a non-trivial task --- not only because exquisite sensitivity is required, but also due to the fact that the position and momentum are dynamically linked. Since Heisenberg's uncertainty principle dictates that a position measurement produces a momentum kick (also known as backaction), this linkage necessarily produces a position disturbance later in the oscillator's cycle, with the resulting noise precluding localisation with precision below the level of the zero-point motion.

%BAE, QND
The most common ways of mitigating this noise --- known as backaction evading (BAE) techniques --- involve making position measurements that are essentially periodic. Using this method, by which the backaction heating only heats the unmeasured quadrature, it follows that an arbitrarily sensitive measurement of the other quadrature is possible. This technique, first theoretically developed by Braginsky et al.\cite{braginsky}, is the prototypical quantum non-demolition (QND) measurement of an oscillator --- since joined by QND protocols for energy\cite{jayich} and atomic spin\cite{sewell}.

%Parametric amp
Periodic modulation of an oscillator's spring constant at twice the resonance frequency, such that one quadrature of motion is amplified, allows one to circumvent practical sensitivity limits that arise due to measurement noise. Accordingly, parametric amplification of this kind has been used in nanomechanical systems\cite{paramp1,paramp3} in addition to microwave systems\cite{micro1} as a way of transducing small signals. Since in principle it adds no extra noise, even in the quantum regime, parametric amplification has long been viewed as a cousin of back-action evasion in this limited sense\cite{rugar}. However, since it influences the measured quadrature of the oscillator, unlike BAE, it cannot be considered a type of QND measurement. 

%Paper summary
Recently, we have proposed\cite{prl1,njp} and demonstrated\cite{prl2} a method to accurately measure and squeeze one quadrature of motion via the orthogonal amplified quadrature by using a detuned parametric drive. In this paper, we show there exists a special case of this detuned mechanical parametric amplification (DMPA), such that which one quadrature is not disturbed by the parametric drive. This quadrature is therefore a QND observable. We show that it is possible to equate a weak measurement of the oscillator to a strong but imperfect backaction evading measurement. This allows us to quantify the effectiveness of DMPA as a QND measurement, and hence directly compare DMPA to one-mode backaction evading protocols as used in cavity optomechanics\cite{clerk,purdy,suh13}, as well as a more recently studied two-mode version\cite{woolley}. We show that the parametric scheme is directly analogous to backaction evading measurement, and that in the regime where the oscillator is localised well below the zero-point motion, the effective measurement strength scales linearly with the parametric drive strength. Hence, conditional quantum squeezing of the mechanical state with weak or inefficient measurement, or in the optomechanical bad cavity regime, is made possible with parametric driving. In addition we show that, in contrast to backaction evasion, approaching the limit of perfect squeezing does not degrade the purity. Furthermore, the purity scales more favourably with the measurement efficiency and is therefore more robust to measurement loss.

%%%%%%%%%%%%%%%%%%%%%%%%%%%%%%%%%%%%%%%%%%%%%%%%%%%%
%%%%%%%%%%%%%%%%%%%%%%%%%%%%%%%%%%%%%%%%%%%%%%%%%%%%
%%%%%%%%%%%%%%%%%%%%%%%%%%%%%%%%%%%%%%%%%%%%%%%%%%%%

\section{Model}

We begin by recapping the essential elements of the DMPA scheme introduced in Refs. \cite{prl1,njp}; unlike the presentation in those works, we focus on the connection to QND measurement.  One starts with a standard optomechanical system where a mechanical resonator of frequency $\omega_m$ is dispersively coupled to a cavity mode of damping rate $\kappa$\cite{kippenberg}.  In the good-cavity limit $\omega_m \gg \kappa$, one can use the cavity output to make a QND measurement of a single mechanical quadrature by simultaneously driving both the red- and blue-detuned mechanical sidebands\cite{clerk}; such a QND measurement when combined with feedback can then in principle lead to high levels of mechanical squeezing.  Here, we focus instead on the bad-cavity limit $\kappa \gg \omega_M$, where the conventional QND scheme for squeezing is impossible: even if one drives both mechanical sidebands, the cavity will couple to (and thus measure) both mechanical quadratures.  

To realize something akin to a QND measurement in the bad-cavity regime, we will take a different approach:  as opposed to engineering the cavity-mechanical interaction by two-tone driving, we will instead modify the coherent dynamics of the mechanical resonator.  This is done by simply introducing a strong parametric modulation of the mechanical spring constant (via, e.g.~electromechanical means\cite{prl1} or otherwise)
at a frequency $2\omega_d$ where $\omega_d = \omega_m + \Delta$. Letting $\hat{x}$ and $\hat{p}$ denote position and momentum, the mechanical Hamiltonian is
\begin{equation}
H = \frac{\hat p^2}{2m} + \frac{\hat x^2}{2}[k_0 + k_r \cos(2t(\omega_m+\Delta))]\; .
\end{equation}
Moving to a rotating frame at the reference frequency $\omega_d$, the position can be decomposed into canonically-conjugate quadratures
\begin{equation}
\sqrt{\frac{m\omega_m}{\hbar}}\,\hat x = \hat X\sin(\omega_d t) + \hat Y\cos(\omega_d t)\; ,
\end{equation}
where $[\hat X,\hat Y] = i$ and the ground state variance is
\begin{equation}
V_g = \langle \hat X^2 \rangle - \langle \hat X \rangle^2 = \langle \hat Y^2 \rangle - \langle \hat Y \rangle^2 = 1/2.
\end{equation}
Introducing creation and annihilation operators in the rotating frame via
\begin{eqnarray}\label{quadx} 
	\hat X = (\hat a + \hat a^\dag)/\sqrt{2}, \,\,\,\,\,  \hat Y = -i(\hat a - \hat a^\dag)/\sqrt{2}. \label{quady}
\end{eqnarray}
and make a rotating-wave approximation (which requires $\omega_m \gg \gamma$, $k_0\gg k_r$), the resulting mechanical Hamiltonian is
\begin{equation}\label{ham}
	\tilde H = \hbar\Delta \hat a^\dag\hat a - \frac{\hbar\chi}{2}(\hat a^2+\hat a^{\dag2}) \; ,
\end{equation}
where $\chi=\omega_m k_r/2 k_0$. The second term causes parametric squeezing of one mechanical quadrature $\hat{U_1} = (\hat X - \hat Y)/\sqrt{2}$ (and amplification of the conjugate quadrature $\hat{U_2} = (\hat X + \hat Y)/\sqrt{2}$) at rate $\chi$, while the first detuning term induces rotation in phase space and hence a mixing of squeezed and amplified quadratures.  At first glance, this additional rotation seems problematic if the eventual goal is mechanical squeezing.  One is thus tempted to set $\Delta = 0$, i.e.~a resonant parametric drive.  In this case, the maximum possible steady-state squeezing of the squeezed quadrature $\hat{U_1}$ is by 50\%, the so-called 3 dB limit\cite{prl1}. For the seemingly ideal case of $\Delta = 0$, this limit cannot be improved by adding continuous position detection\cite{prl1}.

\subsection{Unconditional QND dynamics}

Here, we will instead take a value of the detuning $\Delta$ close to the instability threshold $|\Delta_\mathrm{th}|=\sqrt{\gamma^2+\chi^2}$.  In particular,  one obtains extremely simple dynamics in the case where $\Delta = -\chi$, as the Hamiltonian takes the form
\begin{equation}\label{ham2}
	\tilde H = -\frac{\hbar\chi}{2}\, (\hat X^2+1)\; .
\end{equation}
In this case, the squeezing and rotation operations conspire to produce simple coherent dynamics analogous to that of a free particle:  similar to momentum, the  $\hat{X}$ quadrature is a constant of the motion, while similar to position, $\hat Y$ grows at a rate determined by $\hat X$. That is, in the absence of external influences, $(d/dt) \hat{Y} = \chi \hat{X}$.  It follows trivially that while the $\hat{X}$ quadrature is unaffected by the parametric driving, at long times (or low frequencies) the $\hat{Y}$ quadrature becomes an amplified version of $\hat{X}$. To make this more precise, we include mechanical dissipation in the standard way.  With mechanical amplitude damping at rate $\gamma$,  the quantum Langevin equations describing the mechanical resonator take the form
\begin{equation} \label{lang}
\left[ \begin{array}{c}
\mathrm{d}\hat X \\
\mathrm{d}\hat Y \end{array} \right] = 
\left[ \begin{array}{cc}
-\gamma & 0 \\
\chi & -\gamma \end{array} \right] \, 
\left[ \begin{array}{c}
\hat X \\
\hat Y \end{array} \right] \, \mathrm{d}t  + \sqrt{2\gamma}
\left[ \begin{array}{c}
\mathrm{d}\hat X_\mathrm{in} \\
\mathrm{d}\hat Y_\mathrm{in} \end{array} \right] \; ,
\end{equation}
where $X_\mathrm{in},Y_\mathrm{in}$ describe the input noise from the mechanical bath.  The above is easily solved in the frequency domain as
\begin{eqnarray} \label{spectrum}
\hat X(\omega) = \hat X_0(\omega), \,\,\,\,\,
\hat Y(\omega) = \hat Y_0(\omega) + \frac{2\chi}{\gamma- i\omega}\, \hat X_0(\omega) \; ,
\end{eqnarray}
where
\begin{eqnarray}
\hat X_0(\omega) = \frac{\hat X_\mathrm{in}}{\gamma - i\omega}, \,\,\,\,\,
\hat Y_0(\omega) = \frac{\hat Y_\mathrm{in}}{\gamma - i\omega} \; ,
\end{eqnarray}
are the mechanical quadratures when $\chi = 0$; they simply correspond to the quadratures of a mechanical resonator in thermal equilibrium.

Eqs. \ref{spectrum} express the key idea underlying our DMPA-based backaction evasion scheme:  for low frequencies and large $\chi / \gamma$, the detuned parametric drive causes $\hat{Y}$ to become an amplified version of $\hat{X}$, whereas $\hat{X}$ is completely unaffected by the parametric driving.
The situation is reminiscent of a QND measurement:  the mechanical $\hat Y$ quadrature ``measures" the $\hat X$ quadrature, without any backaction disturbance.  The amplification induced by the detuned parametric driving also means that a standard continuous position measurement made using the cavity output effectively becomes a single-quadrature measurement of $\hat{X}$. We make this precise in what follows.

\subsection{Measurement conditioning}

The next step of the analysis is to understand how a standard conditional position measurement\cite{jacobs} is modified by the effective amplification described above.  As described in  Ref.\ \cite{doherty}, the transformation into a rotating frame allows for simple equations of motion for the conditional state of the oscillator.  As the mechanical Hamiltonian is quadratic and we are making a linear measurement, an initially Gaussian mechanical state will also remain Gaussian at all times.  Hence, the conditional evolution equations reduce to equations for the means and covariance matrix (see \ref{masterapp}).  Letting $V_X \equiv \langle \langle \hat{X}^2 \rangle \rangle$, 
$V_Y \equiv \langle \langle \hat{Y}^2 \rangle \rangle$ and $C \equiv \langle \langle \{ \hat{X} ,\hat{Y}\}/2 \rangle \rangle$ denote the elements of the conditional covariance matrix, one finds
(again taking $\Delta = - \chi$)
\begin{eqnarray}
\label{master1}
\frac{d}{dt}V_X & = & -2\gamma V_X + 2\gamma(N+1/2+N_{BA}) - 4\eta\mu(V_X^2+C^2) \\
\label{mastery}
\frac{d}{dt}V_Y & = & -2\gamma V_Y + 4\chi C + 2\gamma(N+1/2+N_{BA}) - 4\eta\mu(V_Y^2+C^2)\\
\label{master2}
\frac{d}{dt}C & = & -2\gamma C + 2\chi V_X - 4\eta\mu C(V_X+V_Y) \; ,
\end{eqnarray}
where $N$ is the mean bath phonon number, $\mu$ is the measurement rate, and $\eta$ is the efficiency of the measurement ($\eta = 1$ corresponds to a quantum-limited continuous position measurement), while $N_{BA} = \mu / 2 \gamma$ describes the backaction heating of both quadratures from the continuous position measurement, parameterised as an additional mean phonon occupation. These equations have a simple interpretation:  the $\chi$ terms correspond to the coherent dynamics due to the detuned parametric drive, whereas the nonlinear $\mu$-dependent terms correspond to the conditioning terms due to the measurement (i.e.~the measurement leads to an effective nonlinear damping of the variances).  Throughout this paper, we assume a situation where the mechanical susceptibility is not modified by the measurement. In cavity optomechanics, this is achieved by driving the cavity on resonance rather than on the red or blue sidebands.

For comparison, the corresponding conditional evolution equations for a near-QND measurement of the mechanical X quadrature (via, for example, dual sideband driving) take the general form\cite{jacobs,clerk}
\begin{eqnarray}
\label{baemaster1}
\frac{d}{dt}V_X & = & -2\gamma V_X + 2\gamma(N+1/2 + N_{\mathrm{bad}}) - 4\eta\mu V_X^2  \\
\frac{d}{dt}V_Y & = & -2\gamma V_Y + 2\gamma(N+1/2 + N_\mathrm{BA}) - 4\eta\mu C^2 \\
\label{baemaster2}
\frac{d}{dt}C & = & -2\gamma C - 4\eta\mu V_X C  \; ,
\end{eqnarray}
where $N_\mathrm{BA}$ is defined as above, and $N_\mathrm{bad}<N_\mathrm{BA}$ is the spurious backaction heating of the $\hat X$ quadrature due to imperfect QND measurement. Here, the measurement conditioning terms now reflect the fact that only the $\hat X$ quadrature is being measured.  Further, in the ideal QND limit, the parameter $N_{\rm bad} = 0$, and
there is no backaction heating of the measured $\hat X$ quadrature.  Small deviations from the ideal QND limit result in a small amount of backaction heating of the $X$ quadrature; we parameterize this (as in Ref. \cite{clerk}) by $N_{\rm bad}$.  Note that one can easily verify from Eq. \ref{baemaster2} that the stationary conditional state has $C=0$.

\section{QND Measurement via DMPA}

We can now substantiate our claim that as far as the stationary conditional state is concerned, weak measurement of the $\hat Y$ quadrature with strong detuned parametric amplification approximates an efficient QND measurement of $\hat X$. This is done by directly comparing the conditional $\hat X$ quadrature variance equations for DMPA and the backaction evasion case, allowing an effective measurement strength for the $\hat X$ quadrature to be defined. Simple solutions for the $\hat X$ variance and purity are then found in the strong driving limit. Later, the respective roles of the parametric drive and measurement are clarified using the general solution for the effective measurement strength.

\subsection{Effective Measurement Strength}

To define the effective measurement strength, we return to Eqs (\ref{master1}-\ref{master2}) and solve for the stationary value of the covariance $C$.  We obtain
\begin{equation} \label{mueff_imp}
	C = \frac{\chi}{\gamma + 2 \eta \mu (V_X + V_Y) } V_X \equiv g V_X \; .
\end{equation}
Unlike the BAE case, here the stationary value of $C$ is non-zero.  We now use this result to eliminate $C$ from the steady-state equation of motion for $V_X$, obtaining
\begin{equation}\label {qndlike}
	0  =  -2\gamma V_X + 2\gamma(N + 1/2 +N_\mathrm{bad,eff}) - 4\eta\mu_{\rm eff}(V_X,V_Y) V_X^2 \; ,
\end{equation}
where we have described the effective measurement strength:
\begin{equation}
	\mu_{\rm eff}(V_X, V_Y) = \mu \left(1+\frac{C^2}{V_X^2}\right) = \mu (1+g^2) \; ,
\end{equation}
and introduced an effective bad-cavity parameter
\begin{equation}
	N_{\rm bad,eff} = \mu / (2\gamma) \; .
\end{equation}

Comparing against Eq. (\ref{baemaster1}) describing backaction evasion, we see that Eq. (\ref{qndlike}) is now of the same form. For a large measurement enhancement ($g \gg 1$), there is a strong similarity to a near-ideal QND measurement of the $X$ quadrature in that the measurement conditioning parameter $\mu_\mathrm{eff}$ is enhanced far above $\mu$ without a coinciding increase in the backaction heating $N_\mathrm{bad,eff}$, which is independent of $g$.

In the complete absence of measurement ($\mu\!=\!0$), the coherent amplification alone determines the covariance so that $g = C/V_X = \chi'$ (where from here onward $\chi'$ denotes the dimensionless ratio $\chi/\gamma$, equalling unity at the self-oscillation threshold of a non-detuned parametric amplifier).  As the measurement strength is increased, the ratio $g$ is attenuated by the conditioning of the quadratures. This can be seen in Eq.\ (\ref{mueff_imp}), where even though increasing $\mu$ reduces $V_X$ and $V_Y$, the product $\mu(V_X+V_Y)$ is a monotonically increasing function of $\mu$. This attenuation of $g$ reflects the fact that the damping effect of the position measurement on the covariance counteracts the coherent amplification due to the parametric drive. In the limit of a perfect measurement ($\mu/\gamma\rightarrow\infty$), $g$ approaches zero and the parametric drive becomes irrelevant.

A bandwidth picture provides a useful heuristic explanation for the form of Eq.\ (\ref{mueff_imp}), as follows. The measurement conditioning terms $\eta\mu V_X$ and $\eta\mu V_Y$ in this equation also appear in Eqs (\ref{master1}) and (\ref{mastery}), where they may be understood as damping rates in addition to the intrinsic rate $\gamma$. Accordingly, the conditioning associated with a position measurement makes use of information gathered over time-scales $1 /  (\eta\mu V_X)$ and $1 /  (\eta\mu V_Y)$.  However, the effective amplification dynamics are only significant on timescales longer than the mechanical decay time, given by $1/\gamma$. This is shown by Eq.\ (\ref{spectrum}), where the additional term in $\hat Y(\omega)$ decays for $|\omega|\gg\gamma$. Therefore, for sufficiently short measurement timescales the amplification is effectively frozen out and plays no role in the conditioning.\footnote{A rigorous approach to this argument is given in \ref{filterapp} by the filter width parameter used to obtain the optimal position estimate from the measurement results.} This explains the appearance of the rates $\eta\mu V_X$ and $\eta\mu V_Y$ as attenuating terms in the measurement gain given by Eq.\ (\ref{mueff_imp}).

\subsection{Strong Driving Limit}

Since Eq.\ (\ref{mueff_imp}) is an implicit equation, the net effect of the parametric drive and measurement in the regime where the measurement is significant ($2\eta\mu(V_Y+V_X)\gg \gamma$) is not immediately clear. For instance, increasing $\chi'$ will further increase the amplified variance $V_Y$, while increasing the measurement strength $\mu$ will condition $V_Y$ to a smaller value. The situation simplifies in the case of a strong parametric drive ($\chi'\gg 1$), such that the squeezing is strong and $V_Y\gg V_X$.  The net heating of $V_Y$ is then found by keeping only the $\mu V_Y$ term in the denominator of Eq.\ (\ref{mueff_imp}) so that
\begin{equation} \label{capprox}
C\approx\chi V_X/2\eta\mu V_Y \; ,
\end{equation}
and substituting this into Eq. (\ref{mastery}). A cubic equation is then obtained for $V_Y$ in the steady-state, with the solution
\begin{equation}
V_Y \approx \left(\frac{\chi}{\eta\mu}\right)^{2/3}\left(\frac{V_X}{2}\right)^{1/3} \; .
\end{equation}
Inserting this back into Eq. (\ref{capprox}), and then into Eq. (\ref{master1}) using the steady-state condition gives an equation for $V_X$ that can be solved
\begin{equation}\label{vxexp}
0 = -2\gamma V_X + 2\gamma(N+1/2+N_{BA}) - 4\eta\mu V_X^2 - (4\chi^2\eta\mu)^{1/3}V_X^{4/3} \; .
\end{equation}
We can see that the extra conditioning term due to the covariance is now proportional to $V_X^{4/3}$. That is, in the regime where the measurement and parametric drive are both significant, the overall effect of the conditioning via the $Y$ quadrature lies between that of damping (linear in $V_X$) and that of direct conditioning (quadratic in $V_X$).

In the strong driving limit, $V_X$ becomes small enough that the terms proportional to $V_X$ and $V_X^2$  in Eq.~(\ref{vxexp}) can be neglected, yielding the simple solution
\begin{equation} \label{vxsoln1}
V_X \approx \left[\frac{(2N+2N_\mathrm{BA}+1)^3}{4\chi'^2\eta\mu/\gamma}\right]^{1/4} \; .
\end{equation}
Since $N_\mathrm{BA}$ is proportional to the measurement strength $\mu$, there is clearly an optimum value of $\mu$ that minimises $V_X$, located around where this backaction term becomes important. Differentiating to find the optimal measurement strength in this limit yields
\begin{equation}\label{muopt}
\mu_\mathrm{opt}(\chi'\rightarrow \infty) = \gamma(N+1/2) \; ,
\end{equation}
which corresponds to the backaction noise equalling half of the original noise in the oscillator. This trade-off between conditioning and backaction is in contrast to backaction evasion, where the conditional variance of the measured quadrature decreases monotonically with $\mu$, even with spurious heating.

Substituting this optimal measurement strength back into Eq.~(\ref{vxsoln1}) leaves
\begin{equation}\label{vxquant}
V_X \approx \frac{3^{3/4}}{2\eta^{1/4}}\; \sqrt{\frac{2N+1}{\chi'}} \; ,
\end{equation}
Therefore, arbitrarily strong quantum squeezing is possible if $\chi'\gg 2N+1$. This can be compared with the variance obtained for backaction evasion in the strong measurement regime (where $\mu/\gamma\gg 1$)
\begin{equation}
V_X \approx \frac{1}{\eta^{1/2}}\; \sqrt{\frac{2N+1}{\mu/\gamma}} \; .
\end{equation}
Notably, the DMPA scheme is clearly more suited to a sub-optimal efficiency $\eta$, consistent with previous numerical analysis\cite{prl1}. This is especially relevant to nanomechanical systems, where even the best state-of-the-art optomechanical devices have loss factors of the order of 10\%\cite{groblacher}.

\subsection{Squeezing}

\begin{figure}[bh]
\centering
\includegraphics[width=10cm]{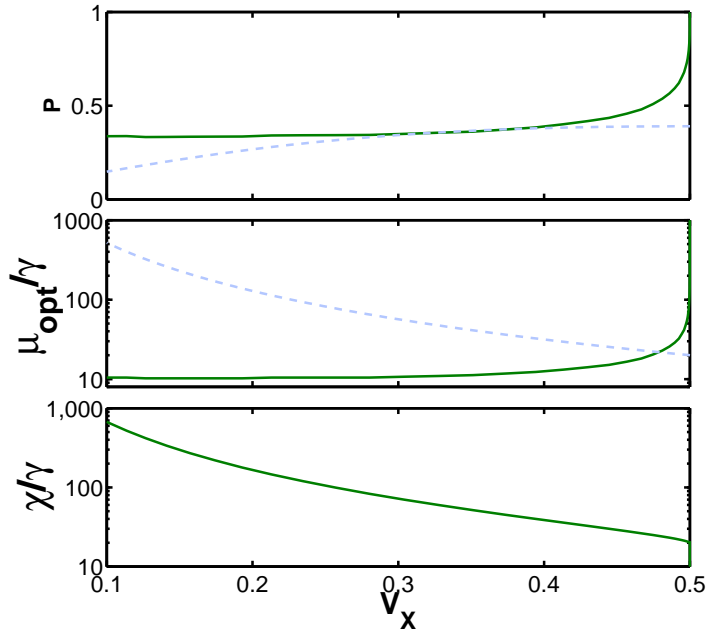}
\caption{\label{purity} Comparison of purity and key parameters for ideal backaction evasion (light, dashed lines) and optimal detuned parametric amplification (solid lines) in the quantum squeezing regime $V_X<0.5$. For DMPA, the measurement strength $\mu$ (middle panel) is optimised to minimise the squeezed variance for each parametric drive strength $\chi/\gamma$ (lower panel). In both cases, the set parameters are an efficiency of $\eta=1$ and the mean thermal phonon occupation of $N=10$.}
\end{figure}

We can now easily see that setting $\mu$ to near the backaction-dominated regime allows the equivalent QND measurement strength $\mu_\mathrm{eff}$ to be deeply within it. Measurement of the proxy $\hat Y$ quadrature can therefore be used to condition the $\hat X$ quadrature to below the level of the zero-point motion. This can be shown in the general case by using numerical solutions to Eqs (\ref{master1}-\ref{master2}). Figure \ref{purity} shows the minimum parametric drive strength required to achieve a fixed level of quantum squeezing using the optimum measurement rate $\mu_\mathrm{opt}$, as well as the purity of the final conditional state. The required measurement strength and achievable purity for backaction evasion are shown for comparison. In the limit of strong squeezing, the parametric drive takes over the measurement's role in backaction evasion, while the optimal measurement strength approaches the constant given by Eq.~(\ref{muopt}) as expected. For low temperatures, this is currently an experimentally feasible parameter, with recent electromechanical\cite{naik} and optomechanical\cite{purdy,suh13} experiments demonstrating backaction noise exceeding zero-point and thermal fluctuations ($\mu > 2\mu_\mathrm{opt}$).

With the measurement strength optimised, the squeezing is limited only by the normalised parametric drive strength $\chi'$. In this analysis, the rotating wave approximation forces the restriction $\chi'\ll Q$, where $Q=\omega_m/\gamma$. Experimental limits on $\chi'$ are also set by the linear response range of the resonator, since the antisqueezed quadrature has variance exceeding the thermal variance by a factor of $1+\chi'^2$. Finally, the condition $\Delta=\pm\chi$ requires precise frequency control of both the resonator and parametric modulation to avoid the instability threshold $|\Delta_\mathrm{th}|=\sqrt{\chi^2+\gamma^2}$, which becomes closer with increasing $\chi'$. Therefore, environmental influences on the oscillator frequency such as temperature fluctuations are detrimental in the strong driving regime, as is also the case for backaction evading protocols\cite{suh12}.

\subsection{Purity}

So far in this analysis, the parallels between DMPA and backaction evasion have been demonstrated for the dynamics and statistics of the $\hat X$ quadrature. It is interesting to note that these parallels do not extend to the orthogonal $\hat Y$ quadrature, which is amplified and conditioned in the DMPA scheme rather than heated. The variance of the $\hat Y$ quadrature is relevant to future quantum applications, many of which rely on a pure or almost-pure squeezed Gaussian state as a building block. These include the production of exotic nonclassical states\cite{neergaard}, entanglement between multiple oscillators\cite{bowen2} and continuous variable quantum computing\cite{datta}. To illustrate the difference between the two schemes considered, we compare the quantity $P=V_g^2/(V_X V_Y - C^2)$, which reaches a maximum value of one for a pure state.

For a backaction evading measurement, the purity can be obtained from the solutions of (\ref{baemaster1}-\ref{baemaster2})
\begin{equation}\label{purity_bae}
P_\mathrm{BAE} = \frac{\eta}{1+\gamma(2N+1)/\mu}\;\frac{2}{\sqrt{1+4\eta\mu(2N+2N_\mathrm{bad}+1)/\gamma} - 1} \; .
\end{equation}
In the ideal good cavity limit $N_\mathrm{bad}=0$ and for a strong measurement $\eta\mu'\gg 2N+1$, the backaction causes a decrease in purity towards zero as $\mu$ is increased
\begin{equation}
P_\mathrm{BAE}(\eta\mu/\gamma \gg 2N+1) \approx \sqrt{\frac{\eta\gamma}{\mu(2N+1)}} \; .
\end{equation}

In contrast to the above, the purity of the steady-state conditional state after applying a detuned parametric drive with the QND condition $|\Delta|=\chi$ and $\mu\neq 0$ can be derived from the general solutions of the variances in Ref.\ \cite{njp}, and written as
\begin{equation}
P_\mathrm{DMPA} = \frac{\eta}{1+\gamma(2N+1)/\mu}\;\left(1 + \frac{2}{\chi'/g - 1}\right) \; .
\end{equation}
In the limit of weak measurement, this purity approaches a very small value due to the amplification of the $Y$ quadrature. However, with an intermediate measurement strength, the conditioning of the $Y$ quadrature allows for a higher purity than the equivalent backaction evading measurement. Since $\chi'/g-1$ is always positive, it is possible to assign a lower bound from the above that is independent of the parametric drive
\begin{equation}
P_\mathrm{DMPA} > \frac{\eta}{1+\gamma(2N+1)/\mu} \; .
\end{equation}
In the strong measurement limit this lower bound on the purity approaches $\eta$, in contrast to Eq.~\ref{purity_bae} where the purity approaches zero for backaction evading measurement. This difference is attributed to the fact that in the DMPA scheme, both quadratures are conditioned by the measurement. Therefore, even though the $\hat Y$ variance is amplified, this quadrature is still kept confined by a nonlinear conditioning term. In contrast, backaction evasion heats the unmeasured $\hat Y$ quadrature, causing $V_Y$ to increase linearly with $\mu'$.

If we consider the optimal measurement strength $\mu_\mathrm{opt}$ that minimises $V_X$ with a fixed parametric drive $\chi$, the purity is reduced from the maximum of $\eta$. This purity is plotted in Figure \ref{purity} for a squeezed $\hat X$ variance (i.e.~$V_X<1/2$), where it is compared with the backaction evading case. It can be clearly seen that while the purity deteriorates as squeezing improves for backaction evasion, the DMPA purity approaches the lower bound of $\eta/3$ (since in this limit $\mu_\mathrm{opt}/\gamma\rightarrow N+1/2$ and $\chi'\gg g$). Furthermore, a compromise can be made by increasing the measurement strength beyond the optimal level, reducing the strength of QND squeezing of the $X$ quadrature in return for higher state purity. This preservation of purity in the strong squeezing limit is in stark contrast to conventional QND quadrature measurement of an oscillator and other methods for steady-state mechanical squeezing. One notable recent proposal using dissipative optomechanics results in purity scaling more favourably than for backaction evasion\cite{kronwald}, however in this case the purity also degrades in the strong squeezing limit.

\subsection{General Solution for Effective Measurement Strength}

Some additional light can be shed on the parallel between DMPA and backaction evasion by quantifying the effective measurement enhancement $\mu_\mathrm{eff}/\mu$. This was found to be equal to $(1+\chi'^2)$ in the limit of no measurement, and reduced to unity in the strong measurement limit. It is between these two limits, where weak measurement and strong parametric driving work in concert, that our scheme finds utility in QND measurement. This intermediate regime --- described above for the limit of strong driving --- will now be examined in detail. Making use of already derived exact solutions to Eqns (\ref{master1}-\ref{master2})\cite{njp}, we can explicitly find $\mu_\mathrm{eff}$ in terms of experimental parameters. This also allows direct comparisons to be made with state-of-the art backaction evasion experiments.

The ratio $\mu_\mathrm{eff}/\mu$, quantifying the ratio of conditioning measurement to backaction-inducing measurement, is given by (see \ref{mueffapp})
\begin{equation} \label{mueffsoln}
\frac{\mu_{\mathrm{eff}}}{\mu} = \frac{2(1+\chi'^2)}{1+\sqrt{(1+4\mathrm{SNR})^2 + 16\chi'^2\mathrm{SNR}} - 4\mathrm{SNR}} \; ,
\end{equation}
where $\mathrm{SNR} = \eta\mu(2N+2N_\mathrm{BA}+1)/\gamma$ defines the signal-to-noise ratio with which the combined thermal and back-action driven motion can be resolved over the measurement noise in the absence of driving. Since $N_\mathrm{BA}\propto \mu$, the inclusion of backaction means that in the limit $N_\mathrm{BA} \gg N+1/2$, the SNR becomes quadratic in $\mu$ rather than linear.

\begin{figure}[bt]
\centering
\includegraphics[width=10cm]{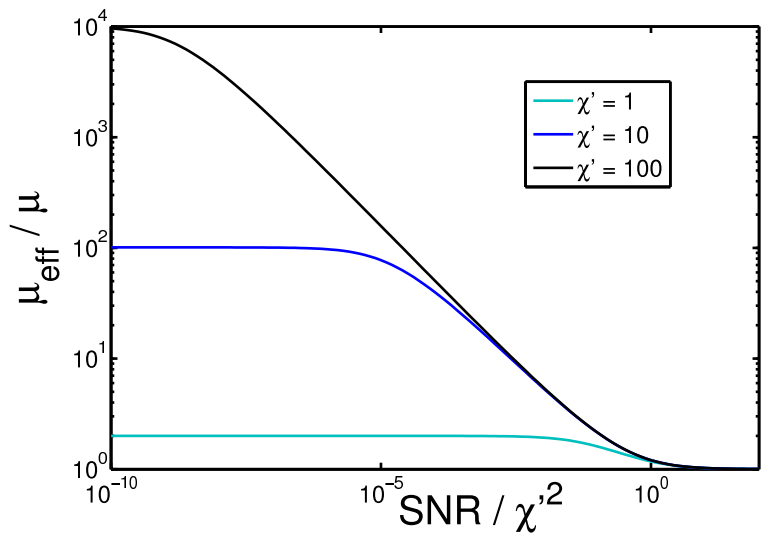}
\caption{\label{mueff} Effective enhancement of the measurement strength as a function of the combined parameter $\mathrm{SNR}/\chi'^2$. For each trace, $\chi'$ is kept constant. The far-left limit corresponds to the weak-measurement limit, where the maximum enhancement is determined by $\chi'$. On the far right, in the strong measurement limit, the enhancement disappears. In the intermediate region, the gradient is approximately $-1/2$, indicating a linear increase with $\chi$.}
\end{figure}

As SNR is increased, the effective measurement enhancement given by Eq.~(\ref{mueffsoln}) passes through three regimes, as illustrated in Figure \ref{mueff} for three values of $\chi'$. For a strong drive ($\chi'\gg 1$), these regimes have simple, well-defined boundaries. The weak measurement limit, where the enhancement is maximised, ends when $\mathrm{SNR}\approx \chi'^{-2}$. Beyond this is an intermediate region of nonzero but reduced gain, where the term $\chi'^2\mathrm{SNR}$ is dominant in the denominator of Eq.~(\ref{mueffsoln}). This corresponds to the amplified $\hat Y$ quadrature being well transduced above the measurement noise. Comparing to Eq.~(\ref{mueff_imp}), this is also where the term $\mu V_Y$ becomes important and the effective measurement ceases to be dominated by the coherent parametric drive. Finally, when SNR exceeds $\chi'^2$, the direct measurement of the $X$ quadrature is more efficient than the proxy measurement and $\mu_\mathrm{eff}/\mu$ approaches 1.

To utilise the full performance of the DMPA-based backaction evasion scheme, the effective measurement strength $\mu_\mathrm{eff}$ must be large compared to $\gamma$, while the spurious heating $N_\mathrm{BA}$ must be weak compared to the thermal noise. It is in the aforementioned intermediate regime that this occurs and the level of quantum squeezing is optimised. When $\chi'\gg 1$ this regime corresponds to an SNR of order unity, signifying that the thermal motion is barely transduced without the aid of the parametric drive. We can then simplify Eq.~(\ref{mueffsoln}) to
\begin{equation}
\frac{\mu_{\mathrm{eff}}}{\mu} \approx \frac{\chi'}{2\sqrt{\mathrm{SNR}}} \; .
\end{equation}
We see that in this intermediate regime, the enhancement is linear with $\chi'$, as is also seen in Figure \ref{mueff}. This linear enhancement is in contrast to the weak measurement limit, where the enhancement scales as $\chi'^2$. Substituting the above expression into Eq.~(\ref{qndlike}) and solving in the limit $\chi' \gg 1$, we get
\begin{equation}
V_X \approx \frac{\mathrm{SNR}^{3/4}}{\sqrt{2\chi'}\eta\mu/\gamma} \; ,
\end{equation}
in exact agreement with Eq.~(\ref{vxsoln1}).

\section{Conclusion}

The goal of backaction evasion is to provide better resolution in a quantum measurement without adding noise to the observable being measured. While parametric amplification is widely known as a means to improve resolution, this generally comes at the cost of disturbing the measured system. In this work, we examine a special case of detuned parametric amplification in which both of these criteria are satisfied; that is, measurement of one quadrature can be enhanced by a parametric drive without disturbing it or increasing the level of backaction noise. We have shown that in the weak measurement regime, the effective enhancement scales as the square of the parametric drive strength, while in the quantum squeezing regime the enhancement scales linearly. Furthermore, in the latter regime, the state purity is favourable compared to traditional backaction evasion, while the squeezing is more robust to measurement inefficiency. With the ability to strongly parametrically drive, this alternative method allows QND measurement in the optomechanical bad-cavity and weak coupling regimes, both of which otherwise preclude this goal. This equivalent approach to backaction evasion is useful for experimental scenarios where purely measurement-based methods may not be viable or sufficient for the preparation of quantum states.

\ack
This research was funded by the Australian Research Council Centre of Excellence CE110001013 and Discovery Project DP0987146. AAC acknowledges support from the DARPA ORCHID program under a grant from the AFOSR. WPB acknowledges DARPA via a grant through the ARO.

\section*{References}

\bibliographystyle{unsrt}
\bibliography{bae1}

\newpage

\appendix

\section{Master equation}
\label{masterapp}

The stochastic master equation, introduced in Ref.\ \cite{prl2}, models the measurement as well as the dissipation of the oscillator, leading to quadrature variances conditioned on the processing of the measurement record. An observer's expected knowledge of the observable $A$ evolves as
\begin{eqnarray}\label{master}
\fl\mathrm{d}\langle\hat A\rangle =& -\frac{i}{\hbar}\langle[\hat A,\tilde H]\rangle\,\mathrm{d}t + 2\gamma\langle\mathcal{D}[\hat a]\hat A\rangle\,\mathrm{d}t + j\sqrt{\eta \mu}\langle\mathcal{H}[\hat X]\hat A\rangle\,\mathrm{d}W_1 + k\sqrt{\eta\mu}\langle\mathcal{H}[\hat Y]\hat A\rangle\,\mathrm{d}W_2 \nonumber \\
\fl & + 2\gamma[N + j(1\!-\!k)N_{\mathrm{bad}} + kN_{\mathrm{BA}}]\langle\mathcal{D}[\hat X]\hat A\rangle\,\mathrm{d}t \\
\fl & + 2\gamma[N + k(1\!-\!j)N_{\mathrm{bad}} + jN_{\mathrm{BA}}]\langle\mathcal{D}[\hat Y]\hat A\rangle\,\mathrm{d}t \nonumber  
\end{eqnarray}
where $N$ is the mean bath phonon number, $\gamma=\omega_m/Q$ is the intrinsic damping rate, $\eta$ is the quantum efficiency and $\mathrm{d}W_1$ and $\mathrm{d}W_2$ are uncorrelated Wiener processes defining the residual noise given the measurement results. The measurement strength $\mu$ defines the maximum amount of conditioning as well as the standard backaction noise via $N_{BA} = \frac{\mu}{2\gamma}$.

The integers $j$ and $k$, where $\{j,k\}\in\{0,1\}$, allow the measurement to be turned off and on in each quadrature. When only one quadrature is measured, the orthogonal quadrature experiences the normal effective increase in phonon occupation due to backaction $N_\mathrm{BA}$. Meanwhile, the measured quadrature experiences a reduced spurious backaction $N_{\mathrm{bad}}$, an amount less than $N_\mathrm{BA}$ and ideally zero. When both quadratures are measured (e.g.\ a continuous position measurement), both experience the full backaction $N_\mathrm{BA}$.

\section{Effective measurement strength}
\label{mueffapp}

The general solution for the conditional variances $V_+,V_-$ when the quadrature phase space is optimally rotated can be found in Ref.\ \cite{njp}. We can use these solutions, including the squeezing angle $\alpha$ to re-obtain $V_X$ and $C$, and thus calculate $g$.
\begin{eqnarray}
V_X &=& \frac{1}{2}(V_+ + V_-) - \frac{1}{2}(V_+ - V_-) \cos(2\alpha) \label{rot1}\\
V_Y &=& \frac{1}{2}(V_+ + V_-) + \frac{1}{2}(V_+ - V_-) \cos(2\alpha) \label{rot2}\\
C &=& \frac{1}{2}(V_+ - V_-) \sin(2\alpha)
\end{eqnarray}
Using the result\cite{njp} that when $\Delta=\chi$
\begin{equation}
V_+ + V_- = \frac{V_+ - V_-}{\cos (2\alpha)}
\end{equation}
we end up, via simple trigonometry, with
\begin{equation}
g = \frac{C}{V_X} = \cot (2\alpha) 
\end{equation}
so
\begin{equation}
\frac{\mu_{\mathrm{eff}}}{\mu} = 1 + g^2 = \frac{1}{\sin^2(2\alpha)}
\end{equation}
An explicit general solution for $\cos (2\alpha)$ is given in Ref.\ \cite{njp}. The effective measurement strength is then easily derived as
\begin{equation}
\frac{\mu_{\mathrm{eff}}}{\mu} = \frac{2(1+\chi'^2)}{1+\sqrt{(1+4\mathrm{SNR})^2 + 16\chi'^2\mathrm{SNR}} - 4\mathrm{SNR}}
\end{equation}
where $\chi'=\chi/\gamma$ is the normalised parametric drive strength and $\mathrm{SNR} = \eta\mu(2N+2N_{BA}+1)/\gamma$ defines the signal-to-noise ratio for the thermal noise.

\section{Filter width}
\label{filterapp}

The relevant timescale of a measurement can be illustrated by the filter parameters --- specifically, the filter width --- that produce the optimal position estimates from the noisy time-series measurements. These parameters are found by Fourier transforming and solving the conditional equations of motion, then transforming back to the time domain\cite{njp}. The exponential decay that specifies the filter width contains the rate
\begin{equation}
\Gamma = \gamma + 2\eta\mu (V_X + V_Y)
\end{equation}
This sum of variances is identical to that for the optimal quadratures (see Eqs \ref{rot1}-\ref{rot2}) and so previously derived results\cite{njp} can be used
\begin{eqnarray}
\Gamma &=& \gamma + 2\eta\mu (V_+ + V_-) \\
   &=& \Delta \tan (2\alpha)
\end{eqnarray}
and using $\Delta=\chi$ in addition to the result $g=\cot (2\alpha)$
\begin{equation}
\frac{\Gamma}{\gamma} = \frac{\chi'}{g}
\end{equation}
The filter width then blows up as $g$ deviates from $\chi'$ and approaches 0. This effect exactly coincides with the enhancement factor $\mu_{\rm eff} / \mu$ dropping from $1 + \chi'^2$ back to $1$, and the amplification becoming redundant. After some algebra, the filter width can be rewritten in terms of experimental parameters as
\begin{equation}
\frac{\Gamma}{\gamma} = \sqrt{(1+4\mathrm{SNR} + \sqrt{(1+4\mathrm{SNR})^2+ 16\chi'^2 \mathrm{SNR}})/2}
\end{equation}
We can see that when $\chi'=0$, the standard expression\cite{njp} is recovered
\begin{equation}
\frac{\Gamma}{\gamma} = \sqrt{1+4\mathrm{SNR}}
\end{equation}
If instead, we let it be non-zero but restrict ourselves to the ultraweak measurement case in which the amplified peak is still obscured under the measurement noise ($\mathrm{SNR}(1+\chi'^2)\ll 1$ where $\chi' \gg 1$), we can expand the inner square root to give
\begin{equation}
\frac{\Gamma}{\gamma} \approx \sqrt{1+4\mathrm{SNR}(1+\chi'^2)}
\end{equation}
which has exactly the same form, but with SNR effectively enhanced by the amplification factor. However, as this approximation breaks down, the filter widens more slowly as a function of this enhanced SNR. In the opposite limit $\mathrm{SNR}(1+\chi'^2)\gg 1$, we obtain
\begin{equation}
\frac{\Gamma}{\gamma} \approx \sqrt[4]{\chi'^2\mathrm{SNR}}
\end{equation}
This is due to the fact that the unconditional Y spectrum only contains a filtered version of the X spectrum (as given by Eq.\ \ref{spectrum}), and hence contains the most accurate X information within a narrow band around the peak. As this peak rises above the noise floor, the measurement fidelity does not scale in the same way as for a direct, high-fidelity measurement of the X quadrature.

\end{document}